\documentclass[11pt,a4paper]{article}
\usepackage{amsmath,amssymb}
\usepackage{epsfig,graphicx}

\begin{document}

\title{\bf Momentum space topology in the lattice gauge theory}

\author{M.A.Zubkov$^a$\footnote{{\bf e-mail}: zubkov@itep.ru}
\\
$^a$ \small{\em ITEP, B.Cheremushkinskaya 25, Moscow, 117259, Russia} 
}
\date{}
\maketitle

\begin{abstract}
Momentum space topology of relativistic gauge theory is considered. The topological 
invariants in momentum space are introduced for the case, when there is the mass gap while the fermion Green functions admit zeros. The index theorem is formulated that relates the number of massless particles and generalized unparticles at the phase transitions to the jumps of the topological invariants. The pattern is illustrated by the lattice model with overlap fermions.  
\end{abstract}

%\textheight 8.9in %s7
%\oddsidemargin -0mm \evensidemargin -0mm \topmargin -1.8cm \textwidth 6.5in

%date{July, 12, 2003}

%\pacs{11.15.Ha, 11.30.-j, 67.90.+z}

% Italic ``theorems''
%\theoremstyle{plain}
\newtheorem{theorem}{Theorem}[section]
\newtheorem{hypothesis}[theorem]{Hypothesis}
\newtheorem{lemma}{Lemma}[section]
\newtheorem{corollary}[lemma]{Corollary}
\newtheorem{proposition}[lemma]{Proposition}
\newtheorem{claim}[lemma]{Claim}

% Roman ``theorems''
%\theoremstyle{definition}
\newtheorem{definition}[lemma]{Definition}
\newtheorem{assumption}{Assumption}

% Humble things: remarks and examples.
%\theoremstyle{remark}
\newtheorem{remark}[lemma]{Remark}
\newtheorem{example}[lemma]{Example}
\newtheorem{problem}[lemma]{Problem}
\newtheorem{exercise}[lemma]{Exercise}

%\newenvironment{nam}[args]{begdef}{enddef}

%\maketitle

\newcommand{\br}{{\bf r}}
\newcommand{\bu}{{\bf \delta}}
\newcommand{\bk}{{\bf k}}
\newcommand{\bq}{{\bf q}}
\def\({\left(}
\def\){\right)}
\def\[{\left[}
\def\]{\right]}

\section{Introduction}

The main subject of the research presented here  is the generalization of topological invariants and index theorem of \cite{VZ2012} to the case when the fermion Green function may contain zeros and nonanalytical exceptional points (see also \cite{Z2012}). We expect that models with such unusual properties may be relevant for the description of the new TeV scale physics expected to appear at the LHC. A unified field theory that works somewhere above TeV, may have the Green functions with zeros, our world may live at the position of the phase transition within a model of this kind. The massless fermions and the generalized unparticles appear at the position of this transition. Their numbers are related to the jumps of the topological invariants across the transition. 

In the vicinity of non - analytical
exceptional point the excitations appear that
 do not look like ordinary particles. We call them generalized unparticles.
This notion is more wide than
the notion of usual fermionic unparticles. Namely, the generalized unparticle
has the propagator with non - analyticity of general form while in
\cite{fermion_unparticle, fermion_unparticle2} the particular forms of the
propagator were derived in accordance with the given scaling dimension $d_U$.

For the general consideration of momentum space topology see also
\cite{NielsenNinomiya1981,So1985,IshikawaMatsuyama1986,Horava2005,Creutz2008,Kaplan2011,Kaplan1992,Golterman1993,Volovik2003}
and \cite{Volovik2011,HasanKane2010,Xiao-LiangQi2011,Wen2012}. 
Below momentum space topology of the fermionic system with zeros of the Green
function and generalized unparticles is illustrated by the example of the
lattice model with overlap fermions \cite{Creutz2011,Overlap,Shrock}. 
There are critical values of mass parameter
$m_0 = 2$, $4$, $6$ and $8$, at which for $m \ne 0$ the topological quantum
phase transitions occur between the insulators with different values of
$\tilde{\cal N}_4$ and $\tilde{\cal N}_5$ (more on topological phase
transitions, at which the topological charge of the vacuum changes while the
symmetry does not, see \cite{Volovik2007}). At these values of $m_0$ the vacuum
states contain the generalized unparticles. At the same time both in these
intermediate states and in the insulator states the zeros of the Green function
are present in momentum space.  The total number of topologically
protected unparticles at the transition point is $n^u_F=\Delta \tilde{\cal N}_4
$.

For the values of $m_0 \ne 0,2,4,6,8$ at the phase transition between the
states with different signs of $m$ there are no unparticles but massless
fermion excitations appear. The number of topologically protected massless
fermions $n_f^0$ is related to the jump of $\tilde{\cal N}_5$: $n_f^0 = \Delta
\tilde{\cal N}_5/2$.

\section{Green functions without zeros and poles}

As well as in \cite{VZ2012} we consider the fermion systems with
the Green function in Euclidean momentum space of the form
 \begin{equation}
  {\cal G} = \frac{Z[p^2]}{g^i[p] \gamma^i -
im[p]}, \quad i = 1,2,3,4 \label{G_}
\end{equation}
Here $Z(p^2)$ is the wave function renormalization function, while $m(p^2)$ is
the effective mass term, $g_a[p]$ are real functions. $Z$, $g$, and $m$ may be
though of as the diagonal matrices if several flavors of Dirac fermions are
present. 

Let us consider the Euclidean Green's function on the 4D lattice ${\cal G}$ as
the inverse Hamiltonian in 4D momentum space and introduce the 5D Green's
function:
\begin{equation}
\label{Green5}
 G^{-1}(p_5,p_4,{\bf p})= p_5 \gamma^5 + {\cal G}^{-1}(p_4,{\bf p}) = (i p_5  + {\cal Q}^{-1}(p_4,{\bf
 p}))(-i \gamma^5)
\,.
\end{equation}
Then one can introduce the topological invariant as  the 5-form  (see also
\cite{Kaplan2011,Kaplan1992,Golterman1993,Volovik2003,SilaevVolovik2010,ZhongWang2010}):
\begin{definition}
\begin{equation}
\label{N_5} {\cal N}_5 = \frac{1}{2 \pi^3 5! i} {\rm Tr}\,  \int G d G^{-1}
\wedge G d G^{-1}\wedge G d G^{-1} \wedge G d G^{-1}\wedge G d G^{-1} \,,
\end{equation}
where the integration is over the  Brillouin zone in 4D momentum space
$(p_4,{\bf p})$ and over the whole $p_5$ axis.
\end{definition}

The properties of this invariant are summarized in the following lemma:
\begin{lemma}
Eq. (\ref{N_5}) defines the topological invariant for the gapped $4D$ system
with momentum space $\cal M$ if the following equation holds:
\begin{equation}
\int_{\partial [{\cal M}\otimes R]}  \, {\bf tr} \left( [\delta {\cal G}^{-1}]
{ G} \,
  d  { G}^{-1}\wedge
  d  { G}\wedge d  { G}^{-1}\wedge
  d  { G}\right)=0,\quad  p_5^2 \rightarrow \infty \label{dN5}
\end{equation}
 This requirement is satisfied, in particular, for the system with compact closed
$\cal M$.

For the system with the Green function of the form (\ref{G_}) expression for
the topological invariant is reduced to
\begin{equation}
\label{N_51} \tilde{\cal N}_5 = \frac{3}{4 \pi^2 4!} \epsilon_{abcde}\, \int
\hat{g}^a\, d \hat{g}^b \wedge d \hat{g}^c \wedge d \hat{g}^d \wedge d
\hat{g}^e,\quad \hat{g}^a = \frac{g^a}{\sqrt{g^cg^c}}
\end{equation}
\end{lemma}

(This lemma corresponds to the Theorem from Sect. 4.3 of \cite{VZ2012}.)

In addition to the invariant $\tilde{\cal N}_5$ let us also consider a
different construction that coincides with  $\tilde{\cal N}_5$ for the case of
free fermions
\begin{definition}
\begin{eqnarray}
\tilde{\cal N}_4 &=& \frac{1}{48 \pi^2} {\rm Tr}\, \gamma^5 \int_{\cal M}
d{\cal G}^{-1}\wedge d {\cal G} \wedge d {\cal G}^{-1} \wedge d {\cal G}
\label{N40}
\end{eqnarray}
Here the integration is over the whole $4D$ space $\cal M$.
\end{definition}

 The expression in
this integral is the full derivative. That's why the given invariant can be
reduced to the integral over the $3D$ hypersurface $\partial {\cal M}$:
\begin{eqnarray}
\tilde{\cal N}_4 & = & \frac{1}{48 \pi^2} {\rm Tr}\, \gamma^5 \int_{\partial
{\cal M}} {\cal G}^{-1} d {\cal G} \wedge d {\cal G}^{-1} \wedge d {\cal
G}\label{N40_}
\end{eqnarray}

 The
last equation is identical to that of for the invariant ${\cal N}_3$ for
massless fermions. Therefore, Eq. (\ref{N40}) defines the topological invariant
if the Green function anticommutes with $\gamma^5$ on the boundary of momentum
space. In particular, for the noninteracting fermions
 $\tilde{\cal N}_4 = {\rm Sp}\, {\bf 1}$ (the number of Dirac
fermions).

The following lemma allows to calculate invariant $\tilde{\cal N}_4$ in general
case:

\begin{lemma}
\label{calcN4} For the Green function of the form (\ref{G_}) with
$\frac{m[p]}{\sqrt{g_a g_a+m^2}} = 0\, (a = 1,2,3,4)$ on $\partial{\cal M}$ we
have:
\begin{equation}
\label{N_41} \tilde{\cal N}_4 = \frac{1}{2 \pi^2 3!} \epsilon_{abcd}\,
\int_{\partial{\cal M}} \hat{g}^a\, d \hat{g}^b \wedge d \hat{g}^c \wedge d
\hat{g}^d,\quad \hat{g}^a = \frac{g^a}{\sqrt{g^cg^c}}
\end{equation}
\end{lemma}
(This Lemma follows when one substitutes Eq. (\ref{G_}) to Eq. (\ref{N40_}).)

\label{N5calc}

Let us introduce the following parametrization
\begin{equation}
\hat{g}_5 = {\rm cos} 2 \alpha, \quad \hat{g}_a = k_a {\rm sin} 2 \alpha
\end{equation}

Vector $k$ may be undefined at the points of momentum space ${\cal M}$, where
$\hat{g}^a = 0, a = 1,2,3,4$. In nondegenerate case this occurs on points $y_i,
i = 1, ...$. Further we call these points the pseudo - poles of the Green
function.
\begin{definition}
The point $y_i$ in momentum space, where $g^a = 0,\, a = 1,2,3,4$ and,
therefore, ${\cal G}^{-1}[p] = m[p]$, is called pseudo - pole of the Green
function.
\end{definition}

Actually, in the majority of cases at these points massive fermion excitations
appear. This is because in these cases (free continuum fermions, lattice Wilson
fermions, overlap fermions, etc) for infinitely small $m[y_i]$ needed to
approach continuum limit, the Green function behaves as
\begin{equation}
{\cal G} \sim \frac{1}{\lambda \,\sum_a (-1)^{n_a} q_a \gamma^a - i m},
\end{equation}
where $q_a = p_a - y_i$,   $\lambda$ is a real constant, $n_a$ are integer
constants. Therefore, in Minkowsky space the usual dispersion relation is
recovered: $E = \sqrt{q^2 + m^2/\lambda^2}$. We formulate this as the following
lemma:
\begin{lemma}
If $g^a \sim \lambda \sum_a (-1)^{n_a} (p^a - y^a)$ while $m\ne 0$ in the small
vicinity of $y \in {\cal M}$, then at this point massive fermion excitation
appears.
\end{lemma}

Let us denote a small vicinity of the pseudo - pole $y_i$ by $\Omega(y_i)$.
According to \cite{VZ2012} we have
\begin{eqnarray}
\tilde{\cal N}_5 &=&  \frac{1}{ \pi^2 4!} \epsilon_{abcd}\, \int_{\sum_{i =
0,1,...}\partial \Omega(y_i)-\partial {\cal M}} (3 \hat{g}_5 - \hat{g}_5^3 )
k^a\, d k^b \wedge d k^c \wedge d k^d \label{N5p}
\end{eqnarray}
 Let us define the $3D$ analogue of the residue.
\begin{definition}
We denote by ${\bf Res}(p)$ the degree of mapping $\{\hat{g}^a: S^3 \rightarrow
S^3\}$:
\begin{eqnarray}
{\bf Res}(p) &=&  \frac{1}{ 2 \pi^2 3!} \epsilon_{abcd}\, \int_{\partial
\Omega(p)} \hat{g}^a\, d \hat{g}^b \wedge d \hat{g}^c \wedge d
\hat{g}^d,\nonumber\\ && p \in \Omega, \quad |\Omega| \rightarrow 0
 \label{CI}
\end{eqnarray}
\end{definition}

Let us also denote
\begin{equation}
{\bf s}(p) = {\rm sign}\, m[p]
\end{equation}

The lemma follows:
\begin{lemma}
\label{calcN5}  Consider the system with the Green function of the form
(\ref{G_}), with $\frac{m[p]}{\sqrt{g_a g_a+m^2}} = 0\, (a = 1,2,3,4)$ on the
boundary of momentum space. Let us denote by $y_i$ the points in momentum
space, where $g^a = 0, \, a = 1,2,3,4$. Then  the topological invariant
$\tilde{\cal N}_5$ is given by
\begin{eqnarray}
\tilde{\cal N}_5 &=&  \sum_{i = 0,1,...} {\bf s}(y_i) {\bf Res}(y_i)
\label{N5f}
\end{eqnarray}
\end{lemma}
(This is Eq. (31) of \cite{VZ2012}.)
For the details of the proofs of the lemmas presented in this section see \cite{VZ2012}.

\section{Green functions with zeros and poles } \label{SectIndTheor}

Here we generalize the definition of the topological invariants to the case, when the Green function may have
 zeros or poles. Construction of invariants $\tilde{\cal N}_4$ and $\tilde{\cal N}_5$ in
 this case requires some care. The correct definition implies that
 first we consider momentum space without some vicinities $\Omega (z_i), \Omega (p_i)$ of the points $z_i$, where
 $\cal G$ has zeros and points $p_i$, where there are poles of the Green function. 
Then all statements of the previous section (proved in \cite{VZ2012}) are valid if we consider the Green function in Momentum space without  $\Omega (z_i), \Omega (p_i)$. That's why, say, in Eq. (\ref{N5p}) and Eq. (\ref{N_41}) of  the previous section we must add $-\sum_i\partial \Omega (z_i) -\sum_j\partial \Omega (p_j)$ to $\partial {\cal M}$. 

 Next, the limit is considered when the sizes of these
 vicinities tend to zero $|\Omega(p_i)|,|\Omega(z_i)| \rightarrow 0$. In order for such a limit to exist the model must
 obey some requirements.

\begin{remark}
In this paper we consider only the cases, when exceptional points of the Green
function are indeed point - like. We do not consider the situation when
exceptional lines or surfaces of the Green function are present. (The case of
the exceptional surface may correspond, in particular, to the Fermi surface.)
\end{remark}

 For example, if the Green function has the form Eq. (\ref{G_}) and $\hat{g}_5 = \frac{m}{\sqrt{g_a g_a+m^2}} = 0\, (a =
 1,2,3,4)$ at $z_i,p_i$, then the boundaries $\partial \Omega(p_i), \partial \Omega(z_i)$ do not
 contribute to the sum in (\ref{N5p}) at $|\Omega(p_i)|, |\Omega(z_i)| \rightarrow 0$. This
 means that the mentioned above limit exists for $\tilde{\cal N}_5$. At the same time under the same
 conditions $\tilde{\cal N}_4$ is the topological invariant at $|\Omega(p_i)|,|\Omega(z_i)| \rightarrow 0$ and
 the points $p_i, z_i$ contribute the sum in Eq. (\ref{N_41}).

In order to make formulas more simple let us also introduce the $3D$ residue at
"infinity":
\begin{eqnarray}
{\bf Res}(\infty) &=&  -\frac{1}{ 2 \pi^2 3!} \epsilon_{abcd}\, \int_{\partial
{\cal M}} \hat{g}^a\, d \hat{g}^b \wedge d \hat{g}^c \wedge d \hat{g}^d
\end{eqnarray}

 Taking into account Lemma \ref{calcN5} and Lemma \ref{calcN4}, we come to the following
\begin{theorem}
\label{N54theorem} Suppose the Green function has the form (\ref{G_}) and at
its poles and zeros as well as on the boundary of momentum space
$\frac{m[p]}{\sqrt{g_a g_a+m^2}} = 0\, (a = 1,2,3,4)$. We denote by $y_i$ the
points, where ${\cal G}^{-1}(y_i) = m(y_i)$ (pseudo - poles of $\cal G$), by
$z_i$ the points, where ${\cal G}(z_i) = 0$, by $p_i$ the points, where ${\cal
G}^{-1}(z_i) = 0$. Then $\tilde{\cal N}_4$ and $\tilde{\cal N}_5$ are well
defined topological invariants. (This means that $\tilde{\cal N}_4$ and $\tilde{\cal N}_5$ are not changed 
under the smooth deformation of $\cal G$ that keeps the listed above conditions.) As a result we have 
\begin{eqnarray}
\tilde{\cal N}_5 &=&  \sum_{i = 0,1,...} {\bf s}(y_i) {\bf Res}(y_i)
\label{N5f__}
\end{eqnarray}
and
\begin{eqnarray}
\tilde{\cal N}_4 &=& - \sum_{i = 0,1,...} {\bf Res}(z_i) -  \sum_{i = 0,1,...}
{\bf Res}(p_i) -  {\bf Res}(\infty)\label{N4f_}
\end{eqnarray}
\end{theorem}

In addition to zeros and poles in general case momentum space may contain non -
analytical exceptional points $q_i$, where $\cal G$ is not defined but, say,
${\cal G}^{-1}$ may differ from zero.
\begin{definition}
The point $q_i$ in momentum space represents generalized unparticle if in its
small vicinity both $\cal G$ and ${\cal G}^{-1}$ are not analytical as
functions of momenta.
\end{definition}

As usual, poles $p_i$ of $\cal G$ (such points that ${\cal G}^{-1}$ is zero at
$p_i$ but remains analytical in its vicinity) represent massless particles.

\begin{remark} If momentum space contains generalized unparticles, then both $\tilde{\cal N}_5$ and
$\tilde{\cal N}_4$ are not well - defined.
\end{remark}

The pattern of the transition from the state at $\beta> \beta_c$ to the state
at $\beta < \beta_c$ can be described in terms of the flow of exceptional
points of $g^a$. Namely, in general there are zeros $y_i$ of $g^a$, where
${\cal G}^{-1} = m$ (we call them pseudo - poles, some of these points become
poles $p_i = y_i$ of ${\cal G}$ if, in addition, $m[y_i] = 0$). Also there are
zeros $z_i$ of ${\cal G}$, where $g^a \rightarrow \infty$. These points cannot
simply disappear when the system is changed smoothly with no phase transition encountered.
 They may annihilate each other if this is allowed by the
momentum space topology. Namely, two zeros $z_i$, $z_j$ may annihilate if ${\bf
Res}(z_i) + {\bf Res}(z_j) = 0$ because in this case they do not contribute the
sum in Eq. (\ref{N4f_}). For the same reason two poles $p_i$, $p_j$ may
annihilate if ${\bf Res}(p_i) + {\bf Res}(p_j) = 0$.  Two pseudo - poles $y_i,
y_j$ may annihilate each other if ${\bf s}(y_i){\bf Res}(y_i) + {\bf
s}(y_i){\bf Res}(y_j) = 0$ because in this case they do not contribute the sum
in Eq. (\ref{N5f__}).

Now we are ready to formulate the generalized index theorem:
\begin{theorem}
\label{indextheorem} Suppose that the $4D$ system with the Green function of
the form (\ref{G_}) depends on parameter $\beta$ and there is a phase
transition at $\beta_c$ with changing of $\tilde{\cal N}_4$ and $\tilde{\cal
N}_5$. At $\beta \ne \beta_c$ the system does not contain generalized
unparticles and massless excitations. $\cal G$ as a function of $\beta$ is
smooth everywhere except for the points, where it is not defined (even at the phase transition). Green
function may have zeros $z_i$, and $m[z_i] \ne \infty$.    Momentum space $\cal
M$ of the $4D$ model is supposed to be either compact and closed or open. In
the latter case we need that $\cal G$ does not depend on $\beta$ on $\partial
{\cal M}$. Then at $\beta = \beta_c$ the number of topologically protected
generalized unparticles is
\begin{equation}
n^u_f  = \Delta \tilde{\cal N}_4 \label{nuz}
\end{equation}
The number of topologically protected massless excitations at $\beta_c$ is
\begin{equation}
n^0_f = \frac{1}{2}\Delta \tilde{\cal N}_5 - \frac{1}{2} \{\sum_{i: z_j
\rightarrow y_i}  {\bf s}(y_i) \, {\bf Res}(y_i)|_{\beta > \beta_c} -\sum_{i:
y_i \rightarrow z_j} {\bf s}(y_j) \, {\bf Res}(y_j)|_{\beta < \beta_c}
\}\label{nfz}
\end{equation}
Here the first sum is over the pseudo - poles that are transformed to zeros at
$\beta_c$ while the second sum is over the zeros that become pseudo - poles.

\end{theorem}

%\begin{proof}

The {\bf Proof}  is given in \cite{Z2012}. 
\begin{remark}
There are two important particular cases:

%\begin{enumerate}

1. If zeros are not transformed to pseudo - poles and vice versa, then
\begin{equation}
n^0_f = \frac{1}{2}\Delta \tilde{\cal N}_5 \label{nfz0}
\end{equation}

2. If ${\rm sign} (m)$ remains constant on $\cal M$ and as a function of
$\beta$, then $n^0_f = 0$ and
\begin{equation}
n^0_f = \frac{1}{2}\Delta \{ \tilde{\cal N}_5 - {\bf s} \tilde{\cal N}_4\} =
0\label{nfz1}
\end{equation}
From here we obtain $\Delta \tilde{\cal N}_5 = {\rm sign}( m) \, \Delta
\tilde{\cal N}_4$ if ${\rm sign}(m) = const$. In the other words,
\begin{eqnarray}
\Delta\{ \sum_{i = 0,1,...} {\bf Res}(z_i) +  \sum_{i = 0,1,...} {\bf
Res}(y_i)\} & = & 0 \label{C0}
\end{eqnarray}

%\end{enumerate}

\end{remark}

\begin{remark}
The total observed number of generalized unparticles and massless fermions may
be larger than $n_f^u$ and $n_f^0$. In this case some of these excitations may
annihilate each other without a phase transition.
\end{remark}

In a similar way the following theorem may be proved

\begin{theorem}
\label{indextheorem2} Suppose that the $4D$ system with the Green function of
the form (\ref{G_}) depends on parameter $\beta$ and there is a phase
transition at $\beta_c$ with changing of $\tilde{\cal N}_4$ and $\tilde{\cal
N}_5$. At $\beta \ne \beta_c$ all excitations are massless. $\cal G$ as a
function of $\beta$ is smooth everywhere except for the points, where it is not
defined. Green function may have zeros $z_i$, and $m[z_i] \ne \infty$. Momentum
space $\cal M$ of the $4D$ model is supposed to be either compact and closed or
open. In the latter case we need that $\cal G$ does not depend on $\beta$ on
$\partial {\cal M}$. Then at $\beta = \beta_c$ the number of topologically
protected generalized unparticles is
\begin{equation}
n^u_f  = \Delta \tilde{\cal N}_4 \label{nuz}
\end{equation}
\end{theorem}

\section{An example: overlap fermions}
\label{lattice}

 In lattice regularization the periodic boundary conditions are used
in space direction and antiperiodic boundary conditions are used in the
imaginary time direction. The momenta to be considered, therefore, also belong
to a lattice:
\begin{equation}
p_a = \frac{2\pi K_a}{N_a}\,  \quad p_4 = \frac{2\pi K_4+\pi}{N_t}, \quad
K_a,K_4 \in Z\quad a = 1,2,3
\end{equation}
Here $N_a, N_t$ are the lattice sizes in $x$, $y$, $z$, and imaginary time
directions, correspondingly. 

For the free Wilson fermions the Green function has the form \cite{Montvay}:

\begin{eqnarray}
{\cal G}& = & \Bigl( \sum_a \gamma_a {\rm sin}\, p_a - i (m + \sum_a (1 - {\rm
cos}\, p_a)) \Bigr)^{-1}\nonumber\\ & = & \frac{ \sum_a \gamma_a {\rm sin}\,
p_a + i (m + \sum_a (1 - {\rm cos}\, p_a)) }{\sum_a {\rm sin}^2\, p_a + (m +
\sum_a (1 - {\rm cos}\, p_a))^2}, \quad a = 1,2,3,4
\end{eqnarray}

\begin{table}
\begin{center}
\begin{small}
\begin{tabular}{|c|c|c|c|c|c|c|c|c|c|c|c|c|c|c|c|c|}
\hline
$m$ & $\tilde{\cal N}_4$ & $\tilde{\cal N}_5$   \\
\hline
$m>0 $ & $0$ & $0$ \\
\hline
$-2 < m < 0$ & $0$ & $-2$  \\
\hline
$-4 < m < -2 $ & $0$ & $6$ \\
\hline
$-6 < m < -4$ & $0$ & $-6$  \\
\hline
$-8 < m < -6 $ & $0$ & $2$  \\
\hline
$ m < -8 $ & $0$  & $0$ \\
\hline
\end{tabular}
\end{small}
\end{center}
\caption{The values of topological invariants $\tilde{\cal N}_4$ and
$\tilde{\cal N}_5$ for free Wilson fermions. } \label{table2}
\end{table}

In Table \ref{table2} the values of $\tilde{\cal N}_5$ for the Wilson fermions
calculated in \cite{VZ2012} are presented. The index theorem states that the
total number $n_F$ of gapless fermions emerging at the critical values of mass
$m$ is determined by the jump in $\tilde{\cal N}_5$:
\begin{equation}
\label{IndexTheorem} n_F=\frac{1}{2}\Delta\tilde{\cal N}_5
\end{equation}
 It is worth mentioning that
$\tilde{\cal N}_4 = 0$ for Wilson fermions.

When the interaction of Wilson fermions with the lattice gauge field ${\cal U}
= e^{i {\cal A}}$ defined on links is turned on, we have (in coordinate space):
\begin{eqnarray}
&&{\cal G}(x,y)  =  \frac{i}{Z} \int  D{\cal U}\, {\rm exp} \Bigl( - S_G[{\cal
U}] \Bigr) \, {\rm Det}  ({\cal D}[{\cal U},m]) {\cal D}_{x,y}^{-1}[{\cal U},m]
\label{Gr}
\end{eqnarray}
where $S_G$ is the gauge field action while
\begin{equation}
{\cal D}_{x,y}[{\cal U},m]  =  - \frac{1}{2}\sum_i [(1 +
\gamma^i)\delta_{x+{\bf e}_i, y} {\cal U}_{x+{\bf e}_i, y}  +  (1 -
\gamma^i)\delta_{x-{\bf e}_i, y} {\cal U}_{x-{\bf e}_i, y}] +  (m + 4)
\delta_{xy}
\end{equation}

Here ${\bf e}_i$ is the unity vector in the $i$ - th direction. Again, the
Green function in momentum space is expected to  have the form \cite{Shrock} of
Eq. (\ref{G_}).

\begin{table}
\begin{center}
\begin{small}
\begin{tabular}{|c|c|c|c|c|c|c|c|c|c|c|c|c|c|c|c|c|}
\hline
$m_0 < 0 $ & -  & -  & -  & -  &  - \\
\hline
$2 > m_0 > 0 $ & $1\otimes m$  & -  & -  & -  &  - \\
\hline
$4 > m_0 > 2$ & $1\otimes m$  & $4\otimes [m(1 - \frac{2}{m_0} )]$ & -  & -  &  - \\
\hline
$6 > m_0 > 4$ & $1\otimes m$  & $4\otimes [m(1 - \frac{2}{m_0} )]$ & $6\otimes [m(1 - \frac{4}{m_0} )]$  & -  &  - \\
\hline
$8 > m_0 > 6$ & $1 \otimes m$  & $4 \otimes [m(1 - \frac{2}{m_0} )]$ & $6\otimes [m(1 - \frac{4}{m_0} )]$  & $4\otimes [m(1 - \frac{6}{m_0} )]$  &  - \\
\hline
$m_0 > 8$ & $1\otimes m$  & $4\otimes [m(1 - \frac{2}{m_0} )]$ & $6\otimes [m(1 - \frac{4}{m_0} )]$  & $4\otimes [m(1 - \frac{6}{m_0} )]$  &  $1\otimes [m(1 - \frac{8}{m_0} )]$ \\
\hline
\end{tabular}
\end{small}
\end{center} \caption{The spectrum of the system with free overlap fermions. In the first column the values of $m_0$ are specified. In
the other columns masses of the doublers are listed.  Expression $x \otimes {v}$ means
$x$ states with the masses equal to $v$. } \label{table5}
\end{table}

Let us consider briefly the properties of overlap fermions. 
In this regularization the propagator has the form:
\begin{eqnarray}
&&{\cal G}(x,y)  =  \frac{1}{Z} \int D D{\cal U}\, {\rm exp} \Bigl( -
\tilde{S}_G[{\cal U}] \Bigr) \, \{-i{\bf D}[{\cal U}] -  i m\}_{x,y}^{-1}
\label{Gr}
\end{eqnarray}
Here the effective action $\tilde{S}_G$ includes also the fermion determinant,
the overlap operator is defined as
\begin{equation}
{\bf D} = \frac{2m_0}{{\cal O}^{-1}-1}
\end{equation}
with
\begin{eqnarray}
{\cal O}_{x,y}[{\cal U}]  & = &  \frac{1}{2}\Bigl( 1 + \frac{{\cal D}[{\cal
U},-m_0]}{\sqrt{{\cal D}^+[{\cal U}, - m_0] {\cal D}[{\cal U},-m_0]}}\Bigr)
\end{eqnarray}
Here $m_0$ and  $m$ are bare mass parameters \cite{Overlap}. The parameter $m$
represents bare physical mass.

In spite of a rather complicated form of the expression for the overlap
operator it is commonly believed that the Green function in momentum space has
the same form (\ref{G_}) as the Green function for Wilson fermions
\cite{Overlap}. In particular, for the free overlap fermions the Green function
has the form:
\begin{equation}
{\cal G}(p) = \frac{1}{g^i[p] \gamma^i - im}
\end{equation}
with (see Appendix in \cite{Overlap}):
\begin{eqnarray}
g^i[p] & = & 2 m_0 \, {\rm sin}\, p^i \, \frac{A(p) + \sqrt{A(p)^2+\sum_i{\rm
sin}^2 \, p^i}}{\sum_i {\rm sin}^2 \,
p^i}, \nonumber\\
A(p) &=& -m_0 + \sum_i [1 -  {\rm cos}\, p^i]
\end{eqnarray}

In Table \ref{table5} we represent the spectrum of the model for different
values of $m_0$. For $0 < m_0 < 2$ one obtains that $\hat{g}^a$ have the only
zero at $p = 0$. However, there are poles of $\hat{g}^a$ at $p_{n_i} = (\pi
n_1, \pi n_2, \pi n_3, \pi n_4), \quad n_i = 0,1, \, \sum n^2 \ne 0$. The
values of $\hat{g}_5$ at these points vanish.

 In general case some of the poles of $\hat{g}^a$ may become zeros depending on the value of $m_0$.
For positive $A(p_{n_i})$ we obtain $g^i[p_{n_i}+\delta p] \sim 4  m_0
\,A(p_{n_i}) (-1)^{n_i} \, \frac{\delta p^i}{|\delta p|^2} $, where $|\delta
p|^2 = \sum_i [\delta p^i]^2 $ and
\begin{equation}
{\cal G}(p_{n_i}+\delta p) \sim \frac{\frac{1}{4  m_0 \,A(p_{n_i})} |\delta
p|^2}{\sum_i (-1)^{n_i} \, \delta p^i \,  \gamma^i - i \frac{m}{4 m_0
\,A(p_{n_i})} |\delta p|^2}\sim \frac{1}{4  m_0 \,A(p_{n_i})} \sum_i (-1)^{n_i}
\, \delta p^i \,  \gamma^i \label{A+}
\end{equation}
We have zeros of the Green function at these points.

For negative  $A(p_{n_i})$ we obtain $g^i[p_{n_i}+\delta p] \sim
\frac{m_0}{|A(p_{n_i})|}  (-1)^{n_i} \, \delta p^i$ and
\begin{equation}
{\cal G}(p_{n_i}+\delta p) \sim \frac{\frac{|A(p_{n_i})|}{ m_0 }}{\sum_i
(-1)^{n_i} \, \delta p^i \,  \gamma^i - i m \frac{|A(p_{n_i})|}{ m_0
}}\label{A-}
\end{equation}
  The values  $A(p_{n_i}) = -m_0 + 2 \sum_i n_i$ at these points are related to
the values of the masses of the doublers: ${\bf m}_{n_i} = m ( 1 -
\frac{2}{m_0} \sum_i n_i ) $.

The special situation appears if $A(p_{n_i}) = 0$ (this occurs for the
intermediate values $m_0 = 2,4,6,8$):
\begin{equation}
{\cal G}(p_{n_i}+\delta p) \sim \frac{\frac{1}{ 2m_0 }}{\sum_i (-1)^{n_i} \,
\frac{\delta p^i}{|\delta p|} \,  \gamma^i - i  \frac{m}{2 m_0 }},\quad |\delta
p| = \sqrt{\sum_i [\delta p^i]^2}\label{A0}
\end{equation}
In this case the Green function is not defined at  $p_{n_i}$ and the
unparticles appear with the propagator equal (up to the normalization constant)
to that of presented in \cite{fermion_unparticle} (follows from Eq. (8) of
\cite{fermion_unparticle} with $\alpha = \beta = 0, \zeta \ne 0, d_U = 2$). At
$m = 0$ we arrive at the propagator given in Eq. (10) of
\cite{fermion_unparticle} with $\alpha = 0, d_U =2$.

It is worth mentioning that for the overlap fermions unlike the Wilson fermions
there are zeros of the Green function at some points in momentum space (see Eq.
(\ref{A+})). Moreover, at the intermediate values of $m_0$ the Green function
is undefined at some points (see Eq. (\ref{A0} ) ). As a result, we need to
consider momentum space without small vicinities of both mentioned types of the
points in order to calculate $\tilde{\cal N}_5$ and $\tilde{\cal N}_4$.
Momentum space, therefore, becomes open. At $m_0 \ne 0,2,4,6,8$ we have at the
points, where $\cal G$ has zeros $\hat{g}_5 = 0$ due to Eq. (\ref{A+}).
Therefore, the conditions of Theorem \ref{N54theorem} are satisfied. For the
intermediate values of $m_0$ this theorem cannot be applied.

The values of $\tilde{\cal N}_4$ and  $\tilde{\cal N}_5$ for overlap fermions
versus parameter $m_0$ are represented in Table \ref{table4}. Let us remind
that unlike the free Wilson fermions at intermediate values of the mass
parameter $m_0 = 0, 2, 4, 6 ,8$ there exist exceptional points in momentum
space such that the Green function is undefined at these points. Due to this
the invariants $\tilde{\cal N}_4$ and $\tilde{\cal N}_5$ are not well defined
in the intermediate states.

When the interaction with the gauge fields is turned on, it is necessary to
check that $\tilde{\cal N}_4$ and $\tilde{\cal N}_5$ remain topological
invariants. Our check shows that both expressions remain the topological invariants.

In the intermediate states at $m_0 = 2,4,6,8, m\ne 0$ there are no true
massless states. Using data of Table \ref{table4} one finds that this is in
accordance with Eq. (\ref{nfz1}) of the index theorem.
 However, there are the generalized unparticles with the Green
function given by Eq. (\ref{A0}) (see the definition in Section
\ref{SectIndTheor}). The number of generalized unparticles is related to the
jump in $\tilde{\cal N}_4$:
\begin{equation}
n_f^u = \Delta \tilde{\cal N}_4
\end{equation}
This relation can easily be checked and is also in accordance with theorem
\ref{indextheorem}.

In the intermediate state with $m = 0, m_0 \ne 0,2,4,5,8$ the generalized
unparticles are absent. Zeros of $\cal G$ remain zeros across the transition.
However, there are massless fermions. Their number is: $0$ for $m_0 < 0$, $1$
for $2 > m_0 > 0$, $5$ for $4
> m_0 > 2$, $11$ for $6 > m_0
> 4$, $15$ for $8 > m_0 > 6$, $16$ for $ m_0 > 8$. At the same time the number
of topologically protected massless fermions given by Eq. (\ref{nfz0}) is:
\begin{equation}
n_f = \frac{1}{2} \Delta \tilde{\cal N}_5
\end{equation}
This  is $0$ for $m_0 < 0$, $1$ for $2
> m_0
> 0$, $-3$ for $4
> m_0
> 2$, $3$ for $6
> m_0 > 4$, $1$ for $8 > m_0 > 6$, $0$ for $ m_0 > 8$. Therefore, except for
the conventional case $2 > m_0 >0$ there are massless fermions in the
intermediate states that are not protected by momentum space topology. When the
interaction with the gauge field is turned on some of them may annihilate each
other so that the total number of massless fermions is reduced without the
phase transition.

There are also mixed intermediate states $m = 0, m_0 = 2,4,6,8$, where both
massless fermions and generalized unparticles are present. The corresponding
transitions satisfy the conditions of  Theorem \ref{indextheorem2}. All pseudo
- poles of the Green function become true poles. The zeros of the Green
function may be transformed to the massless excitations across the transition
points at $m  = 2,4,6,8$. At the corresponding points the generalized
unparticles appear. Their number is equal to the jump of $\tilde{\cal N}_4$.
The corresponding values are listed in Table \ref{table4}.

\begin{table}
\begin{center}
\begin{small}
\begin{tabular}{|c|c|c|c|c|c|c|c|c|c|c|c|c|c|c|c|c|}
\hline
$m_0$ & $\tilde{\cal N}_4$ & $\tilde{\cal N}_5$   \\
\hline
$-m_0 > 0 $ & $0$ & $0$ \\
\hline
$-2 < -m_0 < 0$ & $1$ & ${\rm sign}\, m$  \\
\hline
$-4 < -m_0 < -2 $ & $-3$ & $-3\,{\rm sign}\, m$ \\
\hline
$-6 < -m_0 < -4$ & $3$ & $3\,{\rm sign}\, m$  \\
\hline
$-8 < -m_0 < -6 $ & $-1$ & $-{\rm sign}\, m$  \\
\hline
$ -m_0 < -8 $ & $0$  & $0$ \\
\hline
\end{tabular}
\end{small}
\end{center}
\caption{The values of topological invariants $\tilde{\cal N}_4$ and
$\tilde{\cal N}_5$ for free overlap fermions. } \label{table4}
\end{table}

\section{Discussion}
\label{Conclusions}

The topological invariants do not feel smooth changes of the model. Only a phase transition may lead to the change of the topological invariant. Therefore, if the free system (without gauge fields) and the interacting system (with gauge fields) belong to the same phase, then the values of the topological invariants are the same. So, we may calculate the topological invariant for the free fermions and it will be equal to the same value for the complicated interacting system that is related to the free system by a smooth transformation. 

The vacuum
states of lattice models with fully gapped fermions but with zeros in Green
function (insulating vacua) in 4D space-time are characterized by two
topological invariants, $\tilde{\cal N}_4$ and $\tilde{\cal N}_5$. They are
responsible for the number of generalized unparticles and gapless fermions
which appear at the topological transitions between the massive states with
different topological charges.  

The continuum limit of the model with overlap fermions considered in Section \ref{lattice} at $m = 0, m_0 = 2,4,6,8$ may be
taken seriously. In such limit a continuum theory appears that contains the
unparticle excitations. At the same time, the general properties of the quantum
phase transition with change of $\tilde{\cal N}_4$, $\tilde{\cal N}_5$ can be
applied to the relativistic field theories with fermions.  So, we may have the
new look at the high energy field theoretical models. The entireties, that are
new  for the high energy physics, appear. These are the zeros and the non -
analytical exceptional points of the Green function. We relate the latter
points to the generalized unparticles.

The author kindly acknowledges discussions with G.E.Volovik who initiated this research.  This work was partly supported by RFBR grant 11-02-01227, by the Federal Special-Purpose
Programme 'Cadres' of the Russian Ministry of Science and Education, by Federal
Special-Purpose Programme 07.514.12.4028.


\begin{thebibliography}{15}

\bibitem{VZ2012}
M.A.Zubkov, G.E.Volovik, Momentum space topological invariants for the $4D$
relativistic vacua with mass gap, Nuclear Physics B (2012)
doi:10.1016/j.nuclphysb.2012.03.002, ArXiv:1201.4185

\bibitem{Z2012}
Generalized unparticles, zeros of the Green function, and momentum space topology of the lattice model with overlap fermions,
M. A. Zubkov, 
Phys. Rev. D 86, 034505 

\bibitem{Volovik2007}
G.E. Volovik, Quantum phase transitions from topology in momentum space, in:
Quantum Analogues: From Phase Transitions to Black Holes and Cosmology, Eds.
W.G. Unruh and R. Schutzhold, Springer Lecture Notes in Physics 718/2007, pp.
31-73; cond-mat/0601372.



\bibitem{KuSa} E.Z. Kuchinskii, M.V.
Sadovskii,  Non - Fermi Liquid Behavior in Fluctuating Gap Model: From Pole to
Zero of the Green's function E.Z. Kuchinskii, M.V. Sadovskii JETP 103 (2006)
415, arXiv:cond-mat/0602406




\bibitem{Volovik2011}  G.E. Volovik,
Topology of quantum vacuum, draft for Chapter in proceedings of the Como Summer
School on analogue gravity,
 arXiv:1111.4627.


 \bibitem{EssinGurarie2011}
A.M. Essin and V. Gurarie, Bulk-boundary correspondence of topological
insulators from their Green's functions, Phys. Rev. B {\bf 84}, 125132 (2011).

\bibitem{Gurarie2011}
 V. Gurarie,
 Single-particle Green-s functions and interacting topological insulators,
 Phys. Rev. B {\bf 83}, 085426 (2011).

\bibitem{SilaevVolovik2012}
M.A. Silaev, G.E.Volovik,
 Evolution of edge states in topological superfluids during the quantum phase transition,
 Pis'ma ZhETF {\bf 95}, 29--32 (2012); JETP Lett. {\bf 95},   (2012);
  arXiv:1108.1980.


\bibitem{unparticle}
Howard Georgi (2007). "Another Odd Thing About Unparticle Physics". Physics
Letters B 650 (4): 275-278. arXiv:0704.2457. Bibcode 2007PhLB..650..275G.
doi:10.1016/j.physletb.2007.05.037.

 Howard Georgi (2007). "Unparticle Physics".
Physical Review Letters 98 (22): 221601. arXiv:hep-ph/0703260. Bibcode
2007PhRvL..98v1601G. doi:10.1103/PhysRevLett.98.221601

\bibitem{fermion_unparticle}
Mingxing Luo, Guohuai Zhu, Some Phenomenologies of Unparticle Physics,
Phys.Lett.B659:341-344,2008

\bibitem{fermion_unparticle2}
Rahul Basu, Debajyoti Choudhury, H. S. Mani, Fermionic un-particles, gauge
interactions and the $\beta$ function, Eur.Phys.J.C61:461-470,2009

\bibitem{unparticle_phenomenology}
Kathryn M. Zurek, TASI 2009 Lectures: Searching for Unexpected Physics at the
LHC,  arXiv:1001.2563

 \bibitem{Volovik2010}
 G.E. Volovik,
 Topological invariants  for Standard Model: from semi-metal to topological insulator,
 Pis'ma ZhETF {\bf 91}, 61--67 (2010);   JETP Lett. {\bf 91}, 55--61 (2010);
arXiv:0912.0502.


 \bibitem{NielsenNinomiya1981}
H.B. Nielsen, M. Ninomiya: Absence of neutrinos on a lattice.  I - Proof by
homotopy theory, Nucl. Phys. B \textbf{185}, 20  (1981); Absence of neutrinos
on a lattice. II - Intuitive homotopy proof, Nucl. Phys. B \textbf{193}, 173
(1981).

\bibitem{So1985}
H. So, Induced topological invariants by lattice fermions in odd dimensions,
Prog. Theor. Phys. {\bf 74}, 585--593 (1985).

\bibitem{IshikawaMatsuyama1986}
K. Ishikawa  and T. Matsuyama, Magnetic field induced multi component QED in
three-dimensions and quantum Hall effect, Z. Phys. C {\bf 33}, 41--45 (1986).



\bibitem{Horava2005}
P. Ho\v{r}ava, Stability of Fermi surfaces and $K$-theory, Phys. Rev. Lett.
\textbf{95}, 016405 (2005).

\bibitem{Creutz2008}
M. Creutz, Four-dimensional graphene and chiral fermions, JHEP 04 (2008) 017;
arXiv:0712.1201.


\bibitem{Kaplan2011}
D.B. Kaplan and Sichun Sun, Spacetime as a topological insulator,
arXiv:1112.0302.





\bibitem{Kaplan1992}
 D.B. Kaplan,
Method for simulating chiral fermions on the lattice, Phys. Lett.  B {\bf 288},
342--347 (1992); arXiv:hep-lat/9206013.

\bibitem{Golterman1993}
 M.F.L. Golterman, K.  Jansen and D.B. Kaplan,
Chern-Simons  currents and chiral  fermions on the lattice,
 Phys.Lett. B {\bf 301}, 219--223 (1993):
arXiv: hep-lat/9209003.


 \bibitem{Volovik2003} G.E. Volovik, {\it The Universe in a Helium
Droplet}, Clarendon Press,  Oxford (2003).

\bibitem{HasanKane2010}
M.Z. Hasan and C.L. Kane, Topological Insulators, Rev. Mod. Phys. {\bf 82},
3045--3067 (2010).
% arXiv:1002.3895.

\bibitem{Xiao-LiangQi2011}
Xiao-Liang Qi and Shou-Cheng Zhang, Topological insulators and superconductors,
Rev. Mod. Phys. {\bf 83}, 1057--1110 (2011).


 \bibitem{Wen2012}
Zheng-Cheng Gu, Xiao-Gang Wen, Symmetry-protected topological orders for
interacting fermions -- fermionic topological non-linear sigma-models and a
group super-cohomology theory,
 arXiv:1201.2648.




\bibitem{Creutz2011}
M. Creutz, Confinement, chiral symmetry, and the lattice, arXiv:1103.3304.


\bibitem{Overlap}
Frederic D.R. Bonnet, Patrick O. Bowman, Derek B. Leinweber, Anthony G.
Williams, J. B. Zhang, Overlap Quark Propagator in Landau Gauge,
Phys.Rev.D65:114503,2002

\bibitem{Shrock}
Mario Schrock, The chirally improved quark propagator and restoration of chiral
symmetry, arXiv:1112.5107.

J. B. Zhang, Patrick O. Bowman, Ryan J. Coad, Urs M. Heller, Derek B.
Leinweber, Anthony G. Williams, Quark propagator in Landau and Laplacian gauges
with overlap fermions, Phys.Rev.D71:014501,2005, arXiv:hep-lat/0410045



\bibitem{SilaevVolovik2010}
M.A. Silaev and G.E. Volovik, Topological superfluid $^3$He-B: fermion zero
modes on interfaces and in the vortex core, J. Low Temp. Phys. {\bf 161},
460--473 (2010); arXiv:1005.4672.


\bibitem{ZhongWang2010}
Zhong Wang, Xiao-Liang Qi, Shou-Cheng Zhang, General theory of interacting
topological insulators, arXiv:1004.4229.



\bibitem{Z2011}
M.A.Zubkov,   Fermi point in graphene as a monopole in momentum space, Pis'ma
ZhETF {\bf 95}, 168--174 (2012); arXiv:1112.2474 arXiv:1112.2474


\bibitem{Montvay}
I.Montvay, G.Munster, Quantum fields on a lattice, Cambridge University press,
1994.










\end{thebibliography}
\end{document}